\documentclass[twocolumn,showpacs,prl,aps]{revtex4-1}

\usepackage{dcolumn}
\usepackage{amssymb}

\begin{document}

\title{On the information to work conversion: \\   
a  view from ancient fluctuation-dissipation relations}

\author{Yu. E. Kuzovlev}
\email{kuzovlev@kinetic.ac.donetsk.ua, yuk-137@yandex.ru, kuzovlev@donfti.ru  \\} 
\affiliation{Donetsk Free Statistical Physics Laboratory}


\begin{abstract}
The ``generalized fluctuation-dissipation relations'',
which had anticipated the ``fluctuation theorems'' fifteen years before, 
are applied to presently popular  ``information to useful work conversion'' 
to demonstrate that its success and bounds are determined by a specific ``insider'' 
information rather than formal one. 
\end{abstract} 


\maketitle


\,\, 

{\it 1\,\, Introduction}.\, The subject to be under our  attention in general was interestingly reviewed e.g.  in \cite{iss} 
while a curious example of its individualizing can be found e.g. in \cite{mand}. 
The means for our consideration will be so called 
``generalized fluctuation-dissipation relations'' (FDR) which originally appeared 
in \cite{bk1}-\cite{d} and  for the first time presented the so called ``fluctuation theorems'' 
(FT), as it was explained in \cite{pufn} (see also \cite{apufn}, especially Appendix 
therein, notes in \cite{n1,n2}, and articles  \cite{a1}-\cite{qfr1} 
for various applications and modifications of FDR). 

In all reasonings below we take in mind statistical thermodynamics grounded on 
Hamiltonian mechanics (see introductory words in \cite{n1}), not a 
``stochastic thermodynamics'' mixing principles with uncontrolled conjectures 
(like e.g.  ``Marcovianity'').  Our assumptions concern construction of a system's 
Hamiltonian only but not its evolution (in this respect see \cite{lufn,rs}). 

\,\,\,

\,\,

{\it 2\,\,Hamiltonian }.\, Let us consider a  system with Hamiltonian 
\begin{equation}
H_f(X,\Gamma,Q)\,=\, H_0(X,\Gamma,Q)\,-\,f\cdot Q\,\,, \label{h}
\end{equation}
where \,$\{X,\Gamma,Q\}$\,, 
with \,$X=\{X_1,X_2\,\dots\,X_n\}$\,, 
is full set of system's variables (including that of an inner 
``reservoir'' as well as any inner ``Maxwell demons'') and  $\,f=f(t)\,$ is an external force. 
 
We assume that, first,  the system is thermodynamically indifferent in respect to 
variables \,$X$\, and \,$Q$\, in the sense that  
in absence of the  force they make no influence  on ``partition sum'' over the rest of variables: 
\begin{eqnarray}
q\,\equiv\, \int \exp{\{ -\beta\,H_0(X,\Gamma,Q)\}}\,d\Gamma \,=\,
\texttt{const}\,\, . \label{as1}
\end{eqnarray}
Second, system's phase space is finite in respect to \,$X$\,, 
\begin{eqnarray}
0\,\leq\,X_j\,\leq\, \Delta X \,\, , \label{as2}
\end{eqnarray}
although 
\,$Q$\, can take arbitrary values.  
For instance, all \,$X$\,, and \,$Q$\,  along with them,  may take meaning of some 
``angles'' like similar variables in the model of \cite{mand}. 
So, for certainty, let \,$Q$\, and  \,$X$\,  have positive parities (i.e. do not change signs)  
under time reversal. 

\,\,

\,\,\,

{\it 3\,\,Initial state and processes under interest}.\, 
 Because of (\ref{as1}) the system has no stationary equilibrium probability 
 distribution even if \,$f= 0$\,, not speaking about \,$f\neq 0$\,. 
 Therefore it is  reasonable to consider distributions  
 \begin{eqnarray}
\rho(\cdot \,|\,Q_0)\, \equiv\, \frac {\delta(Q-Q_0)}{\Delta X^n \, q}\, 
e^{ -\beta H_0(\cdot)}\,  \,  \label{dq}
\end{eqnarray}
 with \,$\cdot$\, replacing \,$X, \Gamma,Q$\, and factor \,$\Delta X^n $\, in 
 denominator for normalization to unit. 
 They describe quasi-equilibrium states (ensembles of states) 
 characterized  by fixed instant,  e.g. at time \,$t=0$\,, \,$Q$\,'s value 
 in free unperturbed system (at \,$f=0$)\,. 
  
 Then, at \,$t>0$\,, evolution starting from such initial condition's ensemble   
 but obeying the full perturbed Hamiltonian (\ref{h}) with \,$f\neq 0$\,
 will reveal possible connections between  start values 
 \,$X(t=0) =X_0$\,  and further work 
\begin{eqnarray}
W(t)\,=\, f\cdot (Q(t) -Q(0))  = f\cdot (Q(t) -Q_0)  \,\,  \label{w0}
\end{eqnarray} 
of the external force as switched on at \,$t=0$\,. 
In particular, possibilities to make  quantity (\ref{w0}) artificially negative, - 
and thus convert some amount of reservoir's heat (represented by 
\,$\Gamma$\,) into useful work against the force, - 
due to a ``successful'' special  choice of initial  \,$X(t=0) =X_0$\,. 
  
 Such the ladder set \,$X_0$\, may be thought as a ``program'' for the inner ``demons'' 
 (also represented by \,$\Gamma$\,)  to achieve the success. 
 In other words, program of converting initial information into useful work. 
 Thanks to property (\ref{as1}) establishing of this program in itself 
 meets no  energy cost and destroys  informational, 
 i.e. statistical, equilibrium (uncertainty) only, but not thermodynamical 
 equilibrium. 
 
 \,\,
 
 \,\,
 
 {\it 4\,\,Backward processes an FDR}.\, 
 To  examine this question, consider statistical ensemble average 
  \begin{eqnarray}
  I \equiv \oint \delta(Q(t)-Q_1)\, 
  e^{-\beta W(t)} \,\, \times \nonumber \\ \times \,\, 
  \delta(X(0)-X_0)\, \rho(\cdot|Q_0)\,\,  \label{l0} 
   \end{eqnarray}
 with \,$\oint $\, denoting full phase space  integration.  
 Clearly, in view of (\ref{dq}), it may be written also as 
 \begin{eqnarray}
  I\, =\, \label{l}  \frac 1{\Delta X^n}\, e^{-\beta f \Delta Q} \, 
  P(\Delta Q, t\,|\,X_0,Q_0,f)\,\,  
\end{eqnarray}
 with 
 \[
 \Delta Q\equiv Q(t)-Q(0)\,=\,Q_1-Q_0  
 \]
 and \,$P(\Delta Q, t\,|\,X_0,Q_0,f)$\, 
 being \,$\Delta Q$\,'s probability density distribution (PDD) after time \,$t$\, under the given  initial information 
 about \,$X$\,. 
 
 On the other hand, reverting system's phase trajectories back in time from a given 
moment \,$t>0$\,   to \,$t=0$\,,  recalling that 
  \begin{eqnarray}
W(t)\,=\, H_0(X(t),\Gamma(t),Q(t) )  - H_0(X(0),\Gamma(0),Q(0) )    \,\, \nonumber 
\end{eqnarray} 
 and using general recipes of \cite{bk1}, one can express the same quantity in the form 
  \begin{eqnarray}
  I = \oint \delta(X(t)-X_0)\, \delta(Q(t)-Q_0)\, 
   \rho(\cdot|Q_1)\,\, . \label{r0} 
   \end{eqnarray}
 It is nothing but  
 \begin{eqnarray}
  I\, =\, \label{r}    P(X_0, -\Delta Q, t\,|\,Q_1,f)\,\,  , 
\end{eqnarray}
 where \,$-\Delta Q =Q_0-Q_1$\,  and function   \,$P(X,\Delta Q, t\,|\,Q_0,f)$\, is joint 
 PDD of \,$X$\, and \,$\Delta Q$\, at time \,$t$\, 
 in absence of any initial information about \,$X$\,. 
 
 Thus, we obtain FDR 
 \begin{eqnarray}
   \label{lr}  e^{-\beta f \Delta Q} \, 
  P(\Delta Q, t\,|\,X,Q_0,f)\, \frac 1{\Delta X^n} \, =\, \\ \nonumber =\,  
  P(X, -\Delta Q, t\,|\,Q_0+\Delta Q,f)\,\,  , 
\end{eqnarray}
 which connects ($\ast$) influence of initially stated ``program'' \,$X$\,  onto later 
 \,$\Delta Q$\,'s statistics and ($\ast\ast$)  mutual statistical correlations between 
 simultaneous current \,$X$\,'s and  \,$\Delta Q$\,'s values in the course 
 of ``non-programmed'' evolution. 

\,\,

\,\,

{\it 5\,\,Natural simplifications and  FDR for conditional probabilities}.\, 

As far as the variable \,$Q$\, conjugated with the external force \,$f$\, meets no 
restrictors, its absolute value may be of no importance for an interaction between 
its increments \,$\Delta Q$\, and the ``programming variables'' \,$X$\,. 
Then (\ref{lr}) simplifies to 
\begin{eqnarray}
   \label{ulr}  e^{-\beta f \Delta Q} \, 
  P(\Delta Q, t\,|\,X,f)\, \frac 1{\Delta X^n} \, =\, \\ \nonumber =\,  
  P(X, -\Delta Q, t\,|\,f)\,\,  .  
\end{eqnarray}

Next notice, in view of the properties (\ref{as1})-(\ref{as2}), that the fraction \,$1/\Delta X^n$\, 
in (\ref{ulr})  plays role of eigen marginal PDD of variables \,$X$\,, and integration  
of this FDR over  \,$X$\, reduces it to the native  FDR for the  work in itself \cite{bk1,bk2}, 
\begin{eqnarray}
   \label{lr0}  e^{-\beta f \Delta Q} \, 
  P(\Delta Q, t\,|\,f)  =  
  P( -\Delta Q, t\,|\,f)\,\,  .  
\end{eqnarray}
Division of (\ref{ulr})  by (\ref{lr0}) yields 
\begin{eqnarray}
   \label{clr} \frac   {P(\Delta Q, t\,|\,X,f)} 
   {P(\Delta Q, t\,|\,f)}  =  
  \Delta X^n  \, P(X, t\,|\,-\Delta Q,f)\,\,    
\end{eqnarray}
with function \,$ P(X, t\,|\,\Delta Q,f)$\, meaning conditional PDD of \,$X(t)$\, 
under condition of given increment  \,$\Delta Q$\, during previous time interval \,$(0,t)$\,. 
 
This FDR shows that \,$X$\,-program-induced relative increase (decrease) in probability of the work 
to be \,$W=f\Delta Q$\,  equals to relative increase (decrease) in probability  
of finding just the programming values \,$X(t)=X$\, after spontaneous occurence of 
the opposite-sign work \,$-W= -f\Delta Q$\,. 
Hence, in order to provide (enlarge probability of) 
negative work,  \,$W<0$\,, that is ``information to work conversion'', 
one has to start from those ``good''  \,$X\Rightarrow X(0)$\, whose appearance, 
\,$X(t)\Rightarrow X$\,, is stimulated by opposite-sign   positive work 
(as if the latter was spent on writing useful information). 

\,\, 

\,\,

{\it 6\,\,On statistical averages, inequalities  and estimates}.\, 

Now let us integrate FDR (\ref{ulr}) over \,$\Delta Q$\, to obtain 
\begin{eqnarray}
   \label{xlr}  \left\langle\,e^{ - \beta W(t)} \, | \,X,f\,\right\rangle  \, =\, 
   \Delta X^n\,   P(X, t\,|\,f)\,\, ,  
\end{eqnarray}
where angle brackets on the left designate conditional averaging under fixed   
\,$X(0)=X$\,, while on the right-hand side we see ratio of actual a posteriori 
marginal PDD of \,$X=X(t)$\,  to its a priori uniform (``quasi-equilibrium'') PDD \,$1/\Delta X^n$\,. 
This FDR, in turn,  standardly implies inequality 
\begin{eqnarray}
   \label{in} - \,\beta\, \left\langle\, W(t) \, | \,X,f\,\right\rangle  \, \leq\, 
   \ln{\,(\Delta X^n P(X, t\,|\,f) )} \,\, .   
\end{eqnarray}

In principle, of course, the ratio \,$P(X, t|f)/(1/\Delta X^n)$\, 
may be non-constant (\,$X$\,-dependent), - and thus inevitably different  
from unit, - since generally  \,$P(X, t\,|\,f)$\, represents 
essentially non-equilibrium processes and non-equilibrium system's states. 
Hence, for some good \,$X$\,  choices definitely the logarithm's argument in 
(\ref{in}) is greater than unit. Consequently, for such \,$X$\, this inequality allows 
negative mean value of the work, that is success, at least on average, 
in production of useful work against the force for the expense of reservoir's heat energy. 

At that, (\ref{in}) understandably estimates  upper bound of the success and prompts 
that it can be made arbitrary high, up to values growing \,$\propto t$\,, 
at sufficiently large \,$n$\, (under obvious presumptions about 
system's dynamics and ``kinematics'). 

\,\, 

\,\,

{\it 7\,\,Information and phase volume exchange}.\, 

Leaving more complex statistical averages for separate speculations, let us pay attention 
to one more probabilistic characterization of our ``key'' quantity (\ref{xlr}). 
Namely, merely reformulate (\ref{ulr}) as  
\begin{eqnarray}
   \nonumber   
   \frac {\exp{( -\beta f \Delta Q)} \, 
  P(\Delta Q, t| X(0)=X,f)} {P(-\Delta Q, t| X(t)=X,f)}   =   
  \Delta X^n  P(X, t|f)\,  \, ,  
\end{eqnarray}
where denominator is conditional PDD with condition stated at final point of observation 
time interval instead of its initial point in the nominator. 

Naturally, if \,$X(0)=X$\,  favors, say,  greater \,$\Delta Q$\,, then ,$X(t)=X$\, favors smaller 
\,$-\Delta Q$\,. So, the whole left-side fraction indicates distortion 
of generic time symmetry or asymmetry of \,$\Delta Q$\,'s statistics under influence 
of the conditions. If  the latter  make no real effect, then the fraction must be unit  because of phase 
volume conservation (``Liouville theorem'') in dynamics responsible for \,$\Delta Q$\,. 
Correspondingly, factual effectiveness of the conditions (usefulness of ``programming''   
information) requires significant phase volume exchange between different 
microscopic dynamic channels of transport and fluctuations. In our context here, 
channels related to \,$\Delta Q$\, or to \,$X$\,. 

At that, the Liouville theorem keeps 
on the whole but not for separate channels (subsystems) \cite{bkt,pufn,apufn}. 
This is subject of ``generalized 
fluctuation-dissipation reciprocity relations'' \cite{bk2,bkt,d}. 

\,\, 

\,\,
 
{\it 8\,\,Concluding remarks}.\, 

To resume, we illustrated that the ancient classical  ``generalized fluctuation-dissipation relations'' 
help to recognize time-distributed statistical correlations valuable for 
transforming a part of system's thermal energy into useful work against 
external forces. 

A quantum formulation of this subject 
(in particular, quantum analogues of above FDR) depends on 
one or another rule of mutual ordering  of non-commutative quantum variables 
(operators) and thus a rule of symmetrization of operator exponentials (and other 
functions). Most general quantum FDR not pinned to special orderings were exhaustively 
written out already in \cite{bk1}. 
One interesting variant of the ordering was under our consideration in 
\cite{fdr,qfr,qfr1}. Investigations of other variants and approaches to ``quantum FT'' are 
reflected e.g. in \cite{oth,oth1}. 

Quantum translation of the aforesaid joke reasonings may be placed elsewhere. 
I hope they bring a food for thinking and will not overflow today's sea of 
``fluctuation theorems''.



\end{document}